\documentstyle[12pt,epsfig]{article}

\input{epsf}

\begin{document}

\pagestyle{empty}

\begin{flushright}
{\tt hep-ph/9707276}\\
{CERN-TH/97-153}\\
{DFTT 41/97}\\
{Edinburgh 97/8}\\
{UB-ECM-PF 97/11}\\
\end{flushright}
\vskip 12pt
\begin{center}
\title{}
{\bf WHAT CAN WE LEARN FROM\\
POLARIZED STRUCTURE FUNCTION DATA?} 
\vskip 24pt
Richard~ D.~Ball,$^\star$ Giovanni Ridolfi,$^\ast$\\
Guido~Altarelli$^{\dagger}$ and Stefano Forte$^{\ddagger}$
\vskip12pt
{\em $^\star$Department of Physics and Astronomy,
\footnote{Royal Society University Research Fellow}\\ 
University of Edinburgh, Scotland\\
$^\ast$INFN, Sezione di Genova,\\
Via Dodecaneso 33, I-16146 Genova, Italy\\
$^\dagger$Theoretical Physics Division, CERN,\\ 
CH--1211 Geneva 23, Switzerland \\
$^\dagger$Terza Universit\`a di Roma, Rome, Italy \\
$^\ddagger$INFN, Sezione di Torino,\\
Via P. Giuria 1, I-10125 Torino, Italy\\
$^\ddagger$Departament ECM, Universitat de
Barcelona,\footnote{IBERDROLA visiting professor}\\
Diagonal 647, E-08028 Barcelona, Spain}

\vskip 36pt
\end{center}
{\begin{center}{\bf Abstract}\end{center}
\noindent We summarise the perturbative QCD analysis of the structure 
function data for $g_1$ from longitudinally polarized 
deep inelastic scattering from proton, deuteron and neutron targets, 
with particular emphasis on testing sum rules,  
determining helicity fractions, and extracting the strong coupling
from both scaling violations and the Bjorken sum rule.  
}
\vskip 36pt
\begin{center}
{Talk at {\em Les Rencontres de Physique de la Vallée d'Aoste},\\
La Thuile, March 1997, and at {\em DIS97}, Chicago, April 1997}\\
\smallskip
{\em to be published in the proceedings}\\
\end{center}
\bigskip
\begin{flushleft}
{June 1997}
\end{flushleft}
\eject

\pagenumbering{arabic}
\pagestyle{plain}
\def\toinf#1{\mathrel{\mathop{\sim}\limits_{\scriptscriptstyle
{#1\rightarrow\infty }}}}
\def\tozero#1{\mathrel{\mathop{\sim}\limits_{\scriptscriptstyle
{#1\rightarrow0 }}}}
\def\frac#1#2{{{#1}\over {#2}}}
\def\half{\hbox{${1\over 2}$}}\def\third{\hbox{${1\over 3}$}}
\def\quarter{\hbox{${1\over 4}$}}
\def\smallfrac#1#2{\hbox{${{#1}\over {#2}}$}}
\def\epm#1#2{\hbox{${\lower1pt\hbox{$\scriptstyle +#1$}}
           \atop {\raise1pt\hbox{$\scriptstyle -#2$}}$}}
\catcode`@=11 
\renewcommand\section{\@startsection {section}{1}{\z@}
    {-3.5ex plus -1ex minus -.2ex}{2.3ex plus .2ex}{\bf}}
\renewcommand\subsection{\@startsection {subsection}{1}{\z@}
    {-3.5ex plus -1ex minus -.2ex}{2.3ex plus .2ex}{\it}}
\def\slash#1{\mathord{\mathpalette\c@ncel#1}}
 \def\c@ncel#1#2{\ooalign{$\hfil#1\mkern1mu/\hfil$\crcr$#1#2$}}
\def\lsim{\mathrel{\mathpalette\@versim<}}
\def\gsim{\mathrel{\mathpalette\@versim>}}
 \def\@versim#1#2{\lower0.2ex\vbox{\baselineskip\z@skip\lineskip\z@skip
       \lineskiplimit\z@\ialign{$\m@th#1\hfil##$\crcr#2\crcr\sim\crcr}}}
\catcode`@=12 
\newcommand{\ccaption}[2]{\begin{center}\parbox{0.85\textwidth}
           {\caption[#1]{\small{\it{#2}}}}\end{center}}
\def\GeV{{\rm GeV}}
\def\PR{{\it Phys.~Rev.~}}
\def\PRL{{\it Phys.~Rev.~Lett.~}}
\def\NP{{\it Nucl.~Phys.~}}
\def\NPBPS{{\it Nucl.~Phys.~B (Proc.~Suppl.)~}}
\def\PL{{\it Phys.~Lett.~}}
\def\PRep{{\it Phys.~Rep.~}}
\def\AP{{\it Ann.~Phys.~}}
\def\CMP{{\it Comm.~Math.~Phys.~}}
\def\JMP{{\it Jour.~Math.~Phys.~}}
\def\NC{{\it Nuov.~Cim.~}}
\def\SJNP{{\it Sov.~Jour.~Nucl.~Phys.~}}
\def\SPJETP{{\it Sov.~Phys.~J.E.T.P.~}}
\def\ZP{{\it Zeit.~Phys.~}}
\def\JP{{\it Jour.~Phys.~}}
\def\vol#1{{\bf #1}}\def\vyp#1#2#3{\vol{#1} (#2) #3}
 
The data on the nucleon structure function $g_1$, determined from
measurements of the polarization asymmetry in longitudinally 
polarized deep inelastic scattering, is now sufficient to justify
a sophisticated NLO analysis using the full machinery of perturbative QCD 
\cite{revs}.
It is possible to test sum rules, extract polarized parton distributions 
and determine parton helicity fractions, and even, as we shall see, 
make a reasonably accurate determination of $\alpha_s$. Important issues
in these analyses are the dependence on the functional forms used to 
parameterize the initial parton distributions, and, when estimating 
first moments, 
the implications for the extrapolation to small $x$. Here we 
will discuss the results obtained in a recent analysis \cite{abfr}.

In perturbative QCD structure functions are decomposed into convolutions 
of perturbatively calculable coefficient functions and intrinsically 
nonperturbative parton distribution, which then vary with $Q^2$ 
according to perturbative evolution equations \cite{AP}. For $g_1$ the 
appropriate partons are \cite{BFRa,BFRb}
the nonsinglet and singlet polarized sea quark distributions, 
$\Delta q^{NS}(x,Q^2)$ and $\Delta q^{S}(x,Q^2)$ (where $\Delta q \equiv 
q^\uparrow-q^\downarrow + \bar{q}^\uparrow-\bar{q}^\downarrow$), and the 
polarized gluon distribution 
$\Delta g(x,Q^2)\equiv g^\uparrow-g^\downarrow$. Under
evolution $\Delta q^S$ and $\Delta g$ mix, while $\Delta q^{NS}$ evolves
independently. Valence distributions (of the form $q^\uparrow-q^\downarrow 
-(\bar{q}^\uparrow-\bar{q}^\downarrow)$) also evolve independently, but cannot
be measured in inclusive scattering of virtual photons, and are thus not 
accessible from data on $g_1$ alone (unless of course extra assumptions 
are made about the relative size of valence and sea, as 
in ref.\cite{val,valx}). The coefficient functions $C_q^{NS}$,
$C_q^S$ and $C_g$ and splitting functions $P_{qq}^{NS}$, $P_{qq}^{S}$, 
$P_{qg}$, $P_{gq}$ and $P_{gg}$ are now all available at NLO \cite{NLO}. 

First moments of polarized parton distribution functions are of special 
interest, since they have a direct interpretation as the helicity fractions
of the various constituents. First moments of nonsinglet distributions are
chosen to be independent of $Q^2$ (PCAC), and may then be related to 
nonsinglet axial charges ($g_A\equiv a_3$ and $a_8$ for the $SU(2)$ 
triplet and $SU(3)$ octet) measured in weak baryon decays. The first moment 
of the singlet distribution $\Delta\Sigma\equiv\int_0^1 dx \Delta q^{S}$
may also be chosen to be independent of $Q^2$: this is a direct 
consequence of the Adler-Bardeen theorem. The singlet axial charge 
is then 
\begin{equation}
a_0(Q^2)=\Delta\Sigma - n_f\smallfrac{\alpha_s(Q^2)}{2\pi}\Delta g(Q^2).
\label{azero}
\end{equation}
It evolves multiplicatively, beginning at two loops, because of the 
axial anomaly: this implies that $\Delta g(Q^2)$ grows as $\log Q^2$, 
and that $a_0(\infty)$ and $\Delta\Sigma$ will in general be very different 
\cite{ARCCM}. Changes in factorization scheme away from  
these `AB schemes' \cite{BFRb} result in only small changes in $\Delta g$
(of $O(\alpha_s)$), but large changes in $\Delta\Sigma$ (which however is 
then no longer scale independent).

Sum rules express the first moments of nonsinglet and singlet 
combinations of structure functions $g_1$ in terms of nonsinglet and 
singlet axial charges: 
\begin{equation}
\hbox{$\int_0^1 dx g_1^N(x,Q^2)$} 
= \smallfrac{1}{18}[(1-c^{NS})(3\epsilon a_3+a_8)
+(1-c^{S})4a_0(\infty)],
\label{sumrules}
\end{equation}
where $\epsilon = +1,0,-1$ for N$=$p,d,n respectively. The 
isospin triplet (Bjorken) sum rule, for example, relates the first 
moment of $g_1^p-g_1^n$ to $g_A$. The perturbative corrections 
$c^{NS}$ and $c^{S}$, which begin at NLO (i.e. at $O(\alpha_s)$), are 
now both known at NNNLO (i.e. to $O(\alpha_s^3)$) \cite{NNNLO}.

In order to compute first moments of parton distributions and structure
functions it is necessary to extrapolate the data in the measured region 
down to $x=0$. The behaviour of $F_2^p$ at small $x$ is now well 
understood due to data from HERA \cite{revHera}: while at low $Q^2$ 
$xq^{S}(x,Q^2)$ is fairly flat, in accordance with Regge 
theory since the pomeron intercept is close to unity, as $Q^2$ increases
it rises more and more steeply due to a rise in the gluon distribution 
$xg(x,Q^2)$ driven by gluon bremsstrahlung \cite{das}. By contrast 
Regge theory predicts that $xq^{NS}(x,Q^2)$ decreases rapidly at small $x$
(so $q^{NS}(x,Q^2)\ll q^{S}(x,Q^2)$), though with a rise in 
$q^{NS}(x,Q^2)$ which is then stable under perturbative evolution. 
Now according to Regge theory $|\Delta q^{NS}(x,Q^2)|$ and 
$|\Delta q^{S}(x,Q^2)|$ should be flat (or falling) at small 
$x$ and $Q^2$, and perturbative evolution then leads to an instability 
in both distributions \cite{AR}. However $\Delta q^{S}$ and $\Delta g$ 
mix nontrivially, and the eigenvectors of their evolution are 
such that they eventually have 
opposite sign. As a consequence although $\Delta q^{NS}(x,Q^2)$ and 
$\Delta g(x,Q^2)$ both rise at small $x$ 
and increasing $Q^2$, $\Delta q^{S}(x,Q^2)$ falls steeply and 
$g_1^N(x,Q^2)$ is eventually driven negative \cite{BFRa}. 
Estimates of small $x$ contributions 
using Regge extrapolation alone will then be unreliable, underestimating 
their size and sometimes even giving them the wrong sign.

Here our general philosophy will be to estimate small $x$ contributions 
by performing fits to the data which are consistent with Regge behaviour 
at small $x$ and low $Q^2$, but have the effects of perturbative 
evolution superimposed at higher $Q^2$. Our general procedure is the 
standard one: we take some initial parameterization of 
$\Delta q^{NS}(x,Q_0^2)$, $\Delta q^{S}(x,Q_0^2)$ and $\Delta g(x,Q_0^2)$, 
evolve these three distributions to the $Q^2$ of 
the data \cite{SMC,E143,E142,E154} using NLO perturbative 
evolution (in the AB scheme), 
compute $g_1(x,Q^2)$ using the NLO coefficient functions,
and then adjust the parameters in order to optimize the fit.
The quality of such a fit may be judged from the figure.
In principle it should be possible to fit both triplet and octet
nonsinglet distributions (the charm and bottom distributions 
are generated radiatively from threshold), but in practice this is 
not possible with present data since the triplet dominates
($a_8\ll 3a_3$). We thus assume that both 
distributions have the same shape, and find that the results of the 
fits are very insensitive to the value of $a_8$.

In such fits it is important to assess the dependence of the results
on the form of the initial parameterization, otherwise errors may be 
underestimated. This is particularly true here, where we want to compute
first moments, since the shape of the small $x$ tails may depend
very much on the parameterization adopted. To this end we tried four 
different parameterizations:\\
\indent A: the standard form with powers of $x$ and $1-x$, 
and $Q_0^2=1\GeV^2$;\\
\indent B: powers of $\ln 1/x$ and $Q_0^2=1\GeV^2$;\\  
\indent C: as A, but with $Q_0^2=0.3\GeV^2$ and falling inputs;\\
\indent D: as C, but with valencelike Regge behaviour fixed by hand.\\
All four parameterizations give fits to the data of comparable quality.
Furthermore they all lead to structure functions which eventually 
become negative at small $x$ (see figure), as expected. However the size 
of the small $x$ contribution varies quite widely between the four 
distributions, and we take this spread as an estimate of the error on 
the small $x$ extrapolation. It follows that our first moments for 
$g_1^p$, $g_1^d$ and $g_1^n$ will all be smaller, but with a larger 
error, than those obtained using Regge extrapolations (see table~1).
By contrast our first moment of $g_1^p-g_1^n$ is larger, since 
the nonsinglet distribution rises at small $x$.

Fixing $\alpha_s(m_Z)=0.118\pm 0.005$ \cite{alf}, 
and, very conservatively, $a_8=0.6\pm0.2$, we may use the fits to 
determine $g_A$, $\Delta\Sigma$, $\Delta g$ and $a_0$. The results are 
shown in table~2. The theoretical error includes the uncertainty in 
the small $x$ extrapolation (underestimated in ref.\cite{valx}), 
and the effect of higher order corrections 
estimated by variation of the renormalization and factorization scales
(not included in ref.\cite{valx}). Although the shapes of 
the distributions 
are not fixed very precisely by the data, their first
moments are surprisingly well determined. Fitting $g_A$ is a test 
of the Bjorken sum rule: the value obtained from $\beta$-decay is $1.26$,
so we verify the Bjorken sum rule at the $8\%$ level. Similarly 
fitting $\Delta\Sigma$ may be regarded as a test of the constituent quark 
inspired ansatz
$\Delta\Sigma\approx a_8$: again this is confirmed to within one standard 
deviation (though it must be remembered that such a test is 
only meaningful in AB schemes; in other schemes 
$\Delta\Sigma$ is scale dependent and may take any value one chooses). 
As was predicted in ref.\cite{ARCCM}, a large positive value of 
$\Delta g(1\GeV^2)$ (two standard deviations from zero) is then 
responsible for the small value of the singlet axial charge $a_0$
(compatible results for $\Delta g$ were also found in 
\cite{BFRa,BFRb,val,valx,SMC}). 
Our extrapolation to small $x$ is particularly important here: Regge 
extrapolations give a rather larger value for the axial singlet charge, 
with a smaller (but incorrect) estimate of the error \cite{AlRi,SMC}. There 
is still no totally convincing explanation of why $a_0$ should be 
consistent with zero.

The recently published neutron data now make it possible to include 
$\alpha_s$ in the fit, giving the result
\begin{equation}
\alpha_s(m_Z)=0.120\epm{0.004}{0.005}{\rm (exp.)}
\epm{0.009}{0.006}{\rm (th.)}
\end{equation}
This determination is possible because of the scaling violations
in the nonsinglet channel, which are not sensitive to uncertainties 
in the gluon distribution: without the neutron data the error would
have been twice as large. It is thus not affected by 
uncertainties in the small $x$ extrapolation: the dominant error 
instead comes from 
higher order corrections, which are particularly large as much of the data 
is at fairly low $Q^2$. It is tempting to attempt to reduce this uncertainty 
by determining $\alpha_s$ from the Bjorken sum rule \cite{EK}, exploiting
the NNNLO calculations of ref.\cite{NNNLO}. Unfortunately the  
relatively large error in the Bjorken sum resulting from the 
small $x$ extrapolation makes this method impractical at present: we find 
$\alpha_s(m_Z)=0.118\epm{0.010}{0.024}$.

The test of the Bjorken sum rule, the determination of helicity fractions,
and the result for $\alpha_s$ would all be improved by data 
with better statistics at higher $Q^2$, by an independent measurement 
of $\Delta g$ through open charm production \cite{COMPASS}, or by data at 
smaller $x$ \cite{HERAS}. A better theoretical understanding of polarized 
structure functions at small $x$ would also be useful. However since 
the higher order corrections now involve double logarithms of $x$\cite{KLBER},
rather than the single (and thus factorizable \cite{afsx}) small 
$x$ logarithms in the unpolarized case, this may prove to be a 
considerable challenge.

\vfill\eject

\def\etal{{\it et al.}}

\vfill\eject

\begin{table}[htbp]
\vspace*{1.5cm}
\begin{center}
\begin{tabular} {|l|l|l|l|}
\hline
& $\langle Q^2\rangle$ &Regge extrapolation& pQCD extrapolation \\
\hline
{}~SMC~:~p & 10 & $0.136\pm0.016$ & $0.116\pm 0.022$\\
{}~E143~:~p &  ~3  & $0.127\pm0.010$& $0.107\pm 0.017$\\
\hline
{}~SMC~:~d & 10 & $0.041\pm0.007$ & $0.021\pm 0.016$\\
{}~E143~:~d &  ~3  & $0.042\pm0.010$& $0.021\pm 0.014$\\
\hline
{}~E142~:~n &  ~2   & $-0.033\pm 0.011$& $-0.068\pm 0.015$ \\
{}~E154~:~n &  ~5   & $-0.041\pm 0.007$& $-0.070\pm 0.015$ \\
\hline
\end{tabular}
\caption{Results for complete first moments of $g_1^N$.}
\end{center}
\end{table}

\begin{table}[htbp]
\vspace*{1.5cm}
\begin{center}
\begin{tabular} {|l|llll|}
\hline
& $g_A$ &$\Delta\Sigma$& $\Delta g(1)$& $a_0(\infty)$\\
\hline
Central Value & 1.19 & 0.45 & 1.6 & 0.10 \\
Exp error     & 0.05 & 0.04 & 0.4 & 0.05 \\
Th error      & 0.07 & 0.08 & 0.8 & $\epm{0.17}{0.10}$\\
\hline
\end{tabular}
\caption{Axial charges and helicity fractions.}
\end{center}
\end{table}

\vfill\eject

\begin{figure}[htbp] \centering        
\hbox{\epsfig{file=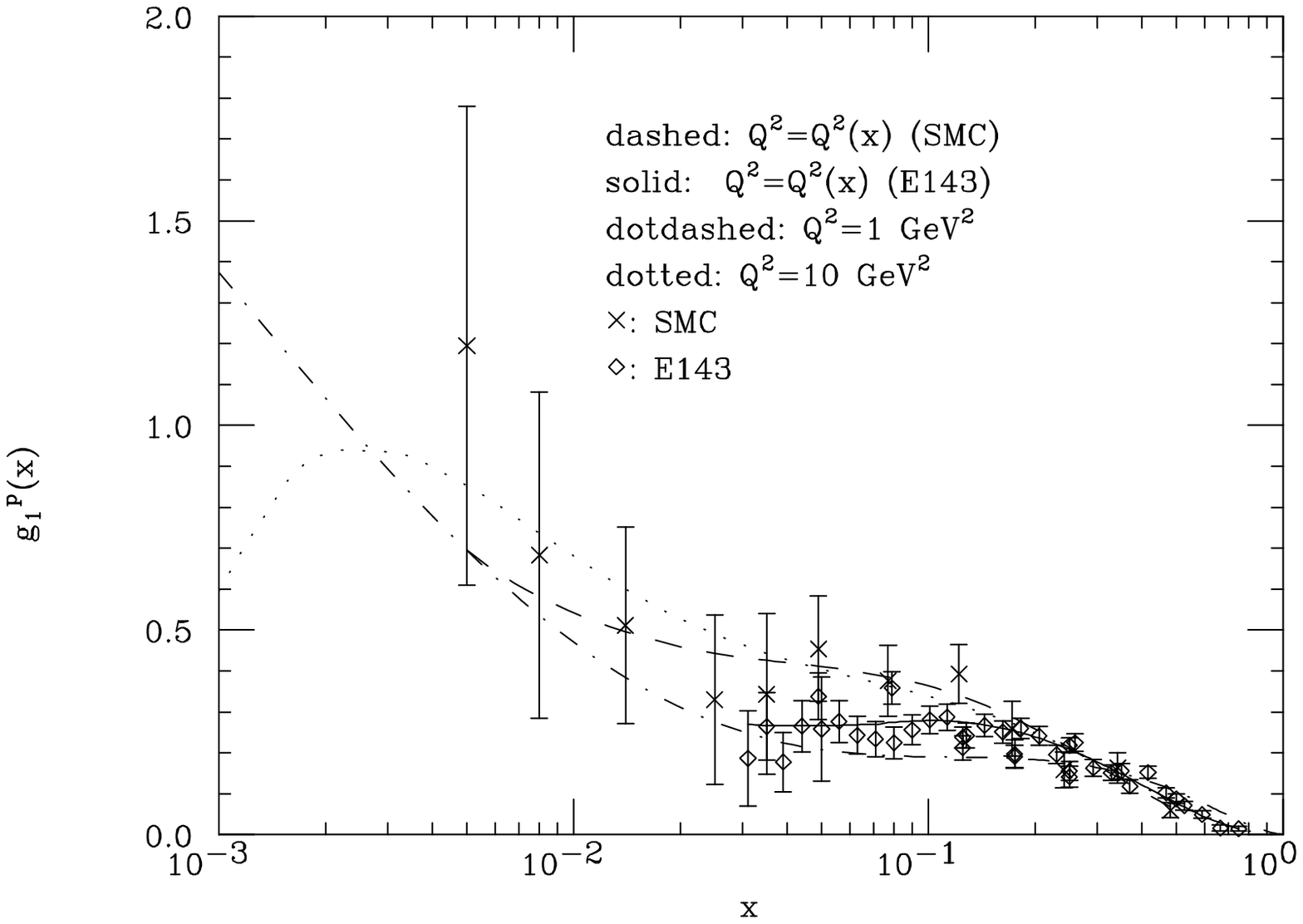,width=74mm}
\epsfig{file=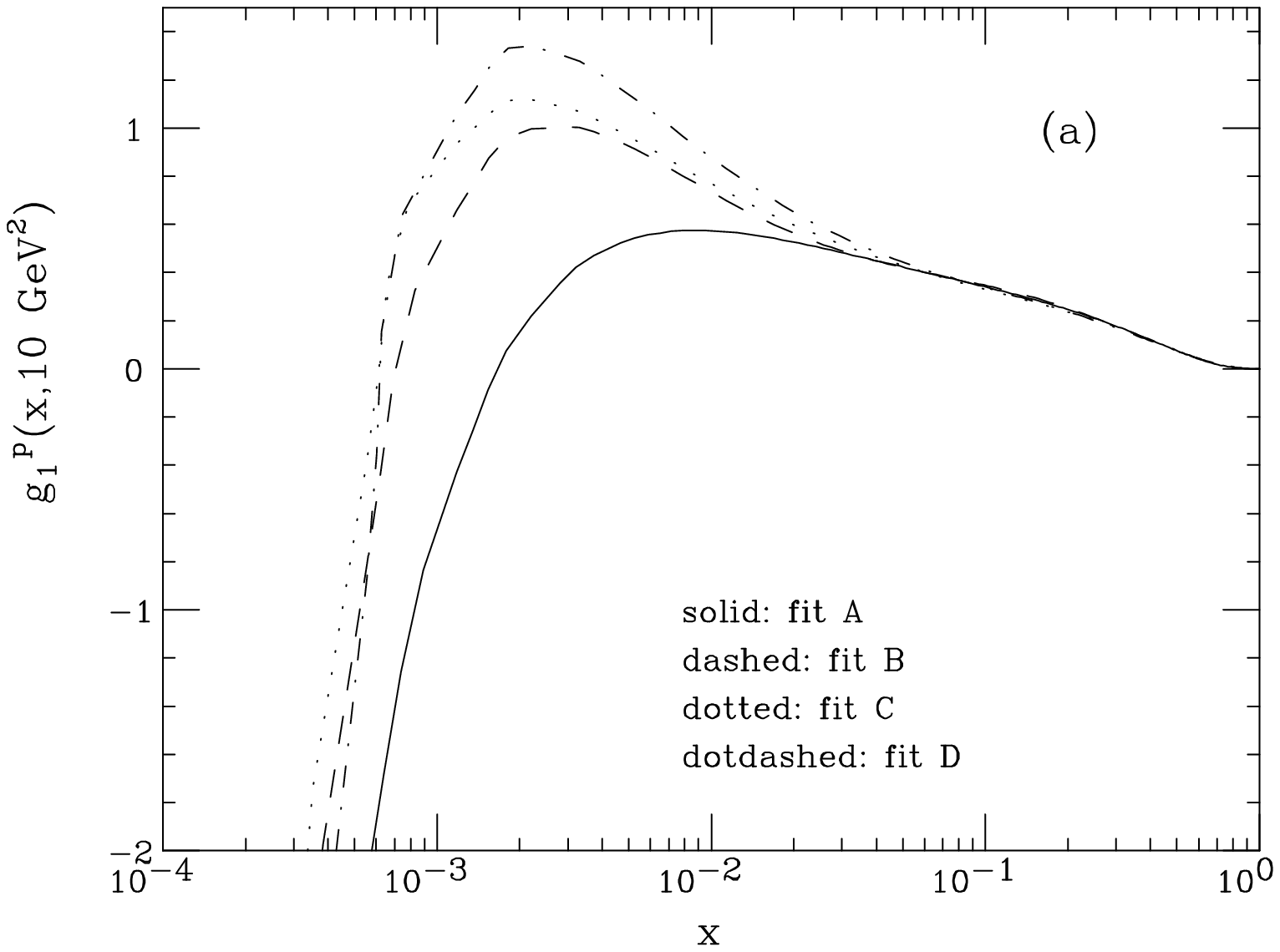,width=70mm}}
\hbox{\epsfig{file=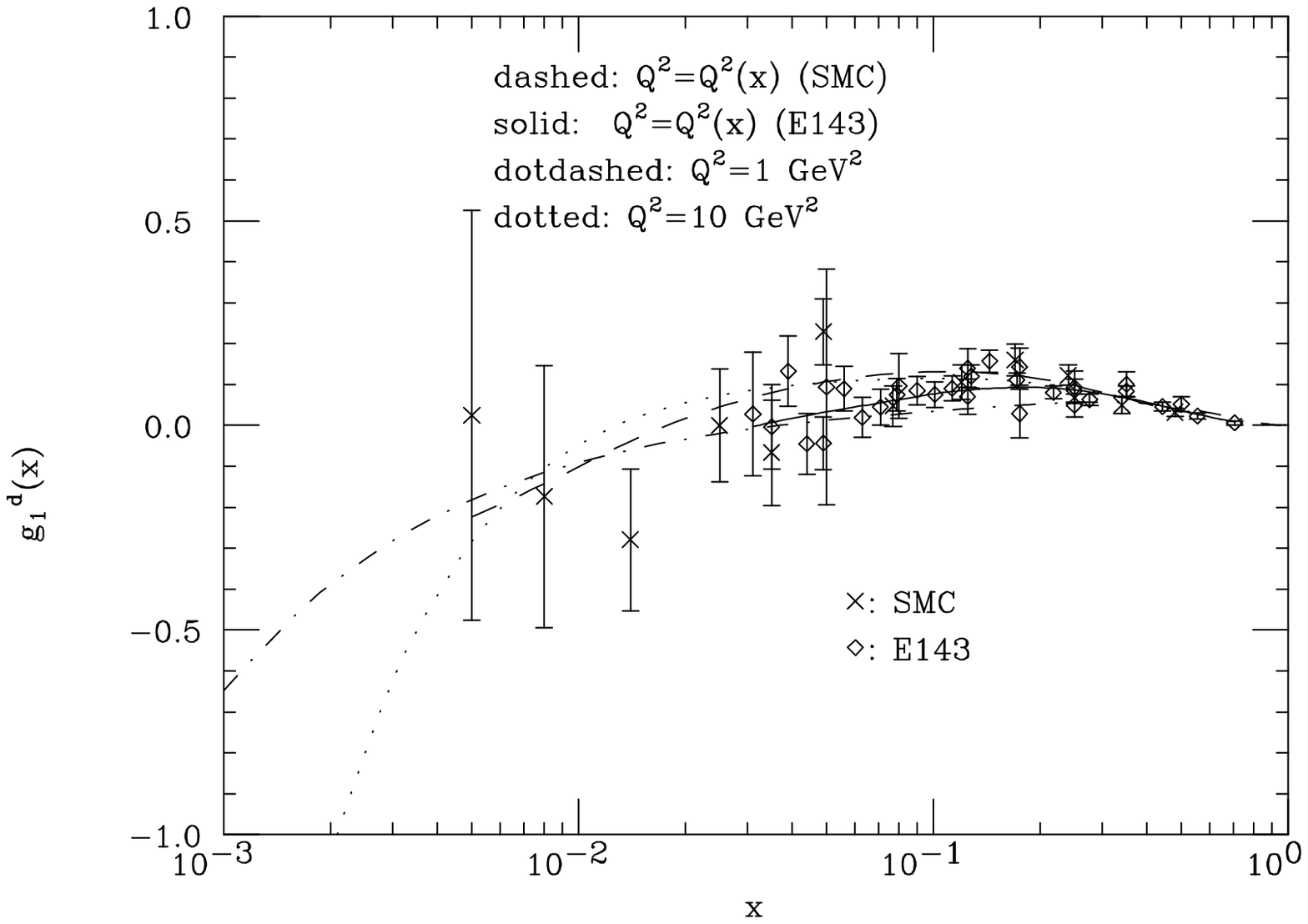,width=74mm}
\epsfig{file=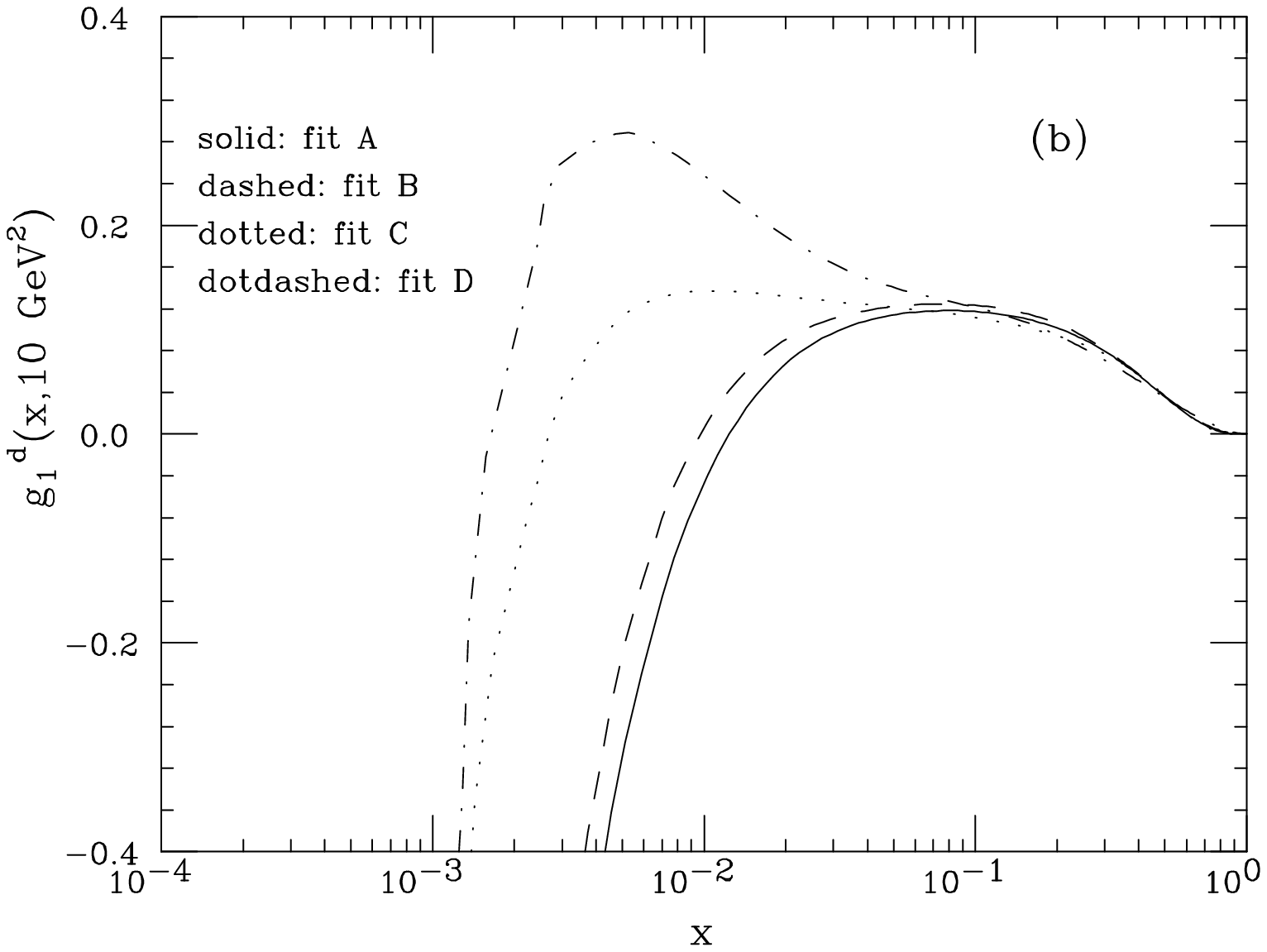,width=70mm}}
\hbox{\epsfig{file=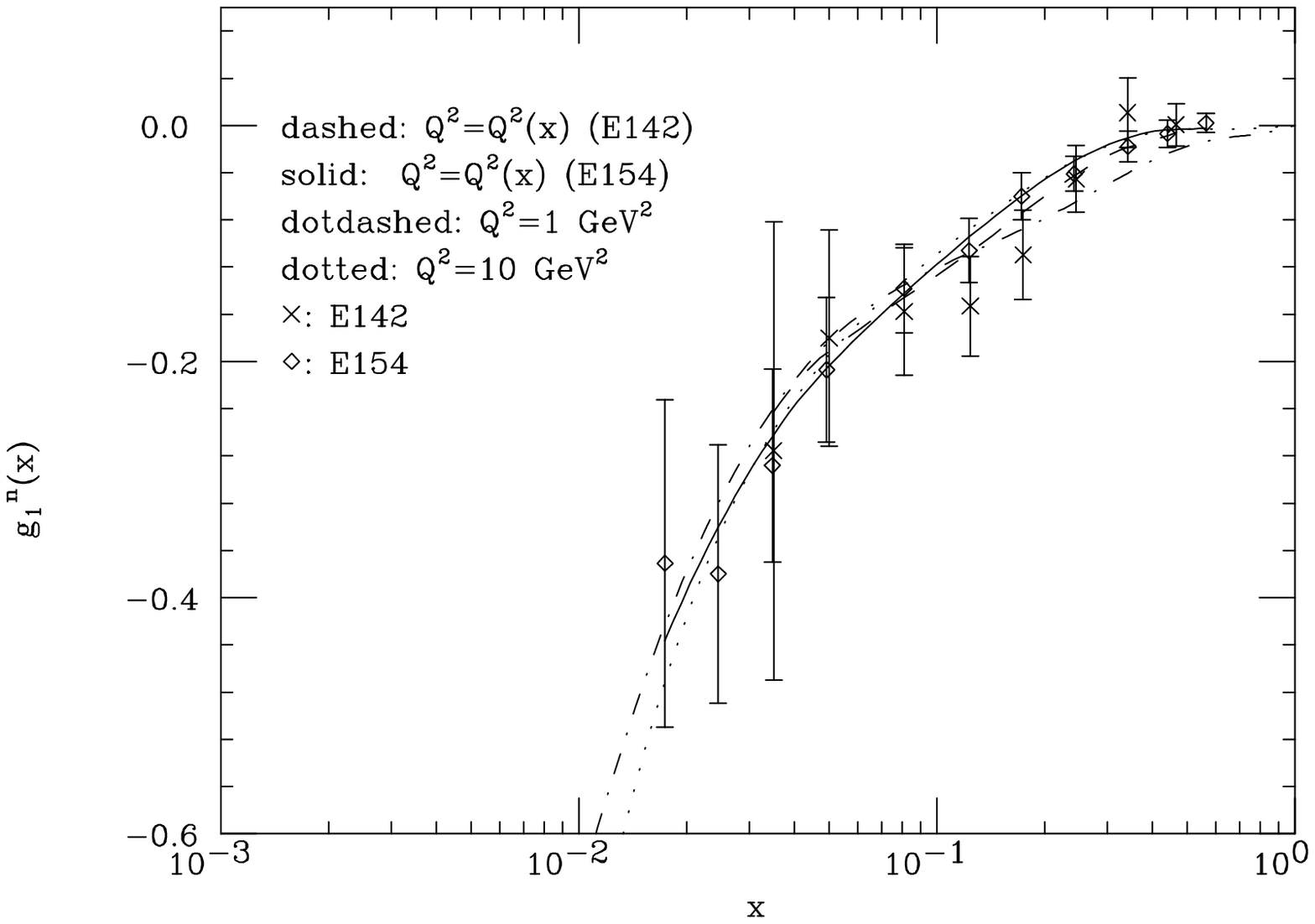,width=74mm}
\epsfig{file=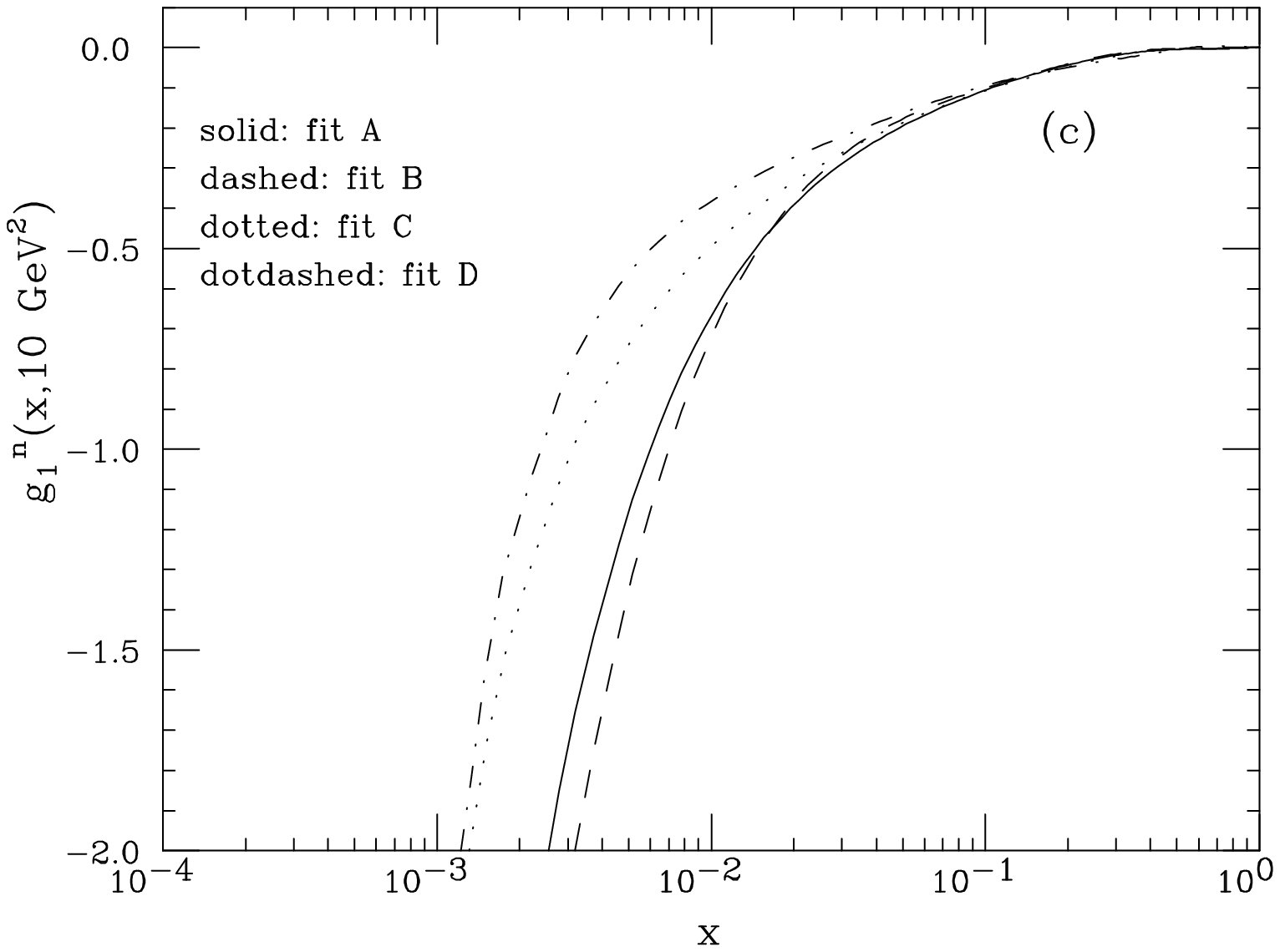,width=70mm}}
\caption[]{Fits to the data: in the left column we plot $g_1^N$, and 
compare fit $B$ to the data, while in the right column we plot $g_1^N$ 
at $Q^2=10\GeV^2$ for fits $A$--$D$, showing the differences between 
the small $x$ tails. Figures 
a,b and c are for proton, deuteron and neutron respectively.}
\label{fig}
\end{figure}                      
 
\end{document}